\documentclass{aastex63} 

\usepackage{amsmath}

\newcommand\SGR{SGR\,1935+2154}

\newcommand\pyne{\url{https://pypi.org/project/pyne2001}}

\received{}
\revised{}
\accepted{}
\submitjournal{ApJL}

\shorttitle{Scintillation towards \SGR{}}
\shortauthors{Simard \& Ravi}

\graphicspath{{./}{figures/}}
\begin{document}

\title{Scintillation can explain the spectral structure of the bright radio burst from \SGR{}}

\correspondingauthor{Dana Simard, Vikram Ravi}
\email{dana.simard@astro.caltech.edu, vikram@caltech.edu}

\author[0000-0002-8873-8784]{Dana Simard}
\affiliation{Cahill Center for Astronomy and Astrophysics, MC 249-17 California Institute of Technology, Pasadena, CA 91125, USA}

\author[0000-0002-7252-5485]{Vikram Ravi}
\affiliation{Cahill Center for Astronomy and Astrophysics, MC 249-17 California Institute of Technology, Pasadena, CA 91125, USA}

\begin{abstract}

The discovery of a fast radio burst (FRB) associated with a magnetar in the Milky Way by the Canadian Hydrogen Intensity Mapping Experiment FRB collaboration (CHIME/FRB) and the Survey for Transient Astronomical Radio Emission 2 (STARE2) has provided an unprecedented opportunity to refine FRB emission models.  The burst discovered by CHIME/FRB shows two components with different spectra.  We explore interstellar scintillation as the origin for this variation in spectral structure.  Modeling a weak scattering screen in the supernova remnant associated with the magnetar, we find that a superluminal apparent transverse velocity of the emission region of $>9.5\,c$ is needed to explain the spectral variation. Alternatively, the two components could have originated from independent emission regions spaced by $>8.3\times10^4$\,km. These scenarios may arise in ``far-away'' models where the emission originates from well beyond the magnetosphere of the magnetar (for example through a synchrotron-maser mechanism set up by an ultra-relativistic radiative shock), but not in ``close-in'' models of emission from within the magnetosphere. If further radio observations of the magnetar confirm scintillation as the source for the observed variation in spectral structure, this scattering model thus constrains the location of the emission region.

\end{abstract}


\section{Introduction} \label{sec:intro}

On the 28th of April, 2020, the fast radio burst project of the Canadian Hydrogen Intensity Mapping Experiment (CHIME/FRB) detected a bright two-component  millisecond-timescale radio burst in the direction of the galactic magnetar \SGR{} \citep{chime_bright_2020}. The coincidence of this burst with a one-second-long hard X-ray burst \citep{mereghetti_integral_2020,ridnaia_peculiar_2020,tavani_xray_2020,zhang_insight_2020} provides an unambiguous association of the CHIME/FRB radio detection with \SGR{}, and stringently constrains models for the radio emission \citep{margalit_implications_2020,lu_unified_2020,yuan_plasmoid_2020,lyutikov_fast_2020}. In addition to the CHIME/FRB-detected burst in the 400--800\,MHz band, a single-component burst was detected by the STARE2 instrument at 1281--1468\,MHz \citep{bochenek_fast_2020}, likely associated with the latter component of the CHIME/FRB detection. A $10^{7}$ times fainter burst was detected two days later by the FAST telescope in the 1.4\,GHz band \citep{zhang_highly_2020}, and \citet{burgay_atel_2020} later reported a tentative detection of persistent faint pulsed emission at 408\,MHz.   

The bright radio burst from \SGR{} exhibited a larger (by a few orders of magnitude) isotropic-equivalent energy than any radio burst previously observed from within the Milky Way. The fluences of $700^{+700}_{-350}$\,kJy\,ms and $1.5\pm0.3$\,MJy\,ms measured by CHIME/FRB and STARE2 respectively, combined with a distance estimate of $\sim 10$\,kpc to \SGR{}, are consistent with a modest low-energy extrapolation of the extragalactic FRB population \citep{bochenek_fast_2020,chime_bright_2020}. This suggests that magnetars like \SGR{} are viable sources of FRBs at extragalactic distances. In interpreting our results in this paper, we focus on two classes of emission models that were developed for extragalactic FRBs, but have been applied to the burst from \SGR{}. ``Close-in'' models \citep{lyutikov_fast_2020,lu_unified_2020} posit that the radio emission is generated at heights of $\lesssim100\,r_\mathrm{NS}$, where $r_\mathrm{NS}\sim10$\,km is the neutron-star (NS) radius. A possible mechanism is the decay, at $\sim20\,r_\mathrm{NS}$, of Alfv\'en waves launched from the magnetar surface by the same disturbance that generates the hard X-ray emission \citep{lu_unified_2020}. On the other hand, ``far-away'' models require that magnetar bursts cause the ejection of a portion of the magnetosphere (known as a plasmoid) at relativistic speeds into the surrounding medium \citep{margalit_implications_2020,yuan_plasmoid_2020}.  The plasmoid will shock the highly magnetized surrounding medium at distances of $10^{11}$\,cm \citep{margalit_implications_2020} to $10^{13}$\,cm \citep{yuan_plasmoid_2020} with Lorentz factors of a few tens \citep{margalit_implications_2020} to a few hundreds \citep{yuan_plasmoid_2020} depending on the composition of the medium. The shock will primarily dissipate radiatively in the high-energy portion of the electromagnetic spectrum, but will also be accompanied by prompt radio emission through the synchrotron maser mechanism. The position and velocity of the radio-emitting region are thus important discriminants between different models for the bright radio burst from \SGR{}.

The two components in the burst detected by CHIME/FRB have different spectra.  The earlier burst is brightest in the bottom of the observing band (400--600\,MHz) while the later burst is faint at frequencies below $\sim 500$\,MHz, but otherwise occupies the majority of the 400--800\,MHz band (see Fig.\ 1 of \citet{chime_bright_2020}.) Both components appear to be temporally broadened by the effects of multi-path propagation through the Milky Way interstellar medium (ISM), commonly referred to as ``scattering''. \citet{chime_bright_2020} fit a characteristic scattering timescale of $0.759\pm0.008$\,ms simultaneously to the two components at 600\,MHz. Scattering was also tentatively observed in the higher-frequency burst detected by STARE2 \citep{bochenek_fast_2020}. 

Spectral differences in emission components have been seen from extragalactic repeating FRBs \citep{hessels_frb_2019,chime_second_2019,chime_discovery_2019}; however, in these repeating FRBs the peak frequencies of components that arrive later are invariably lower, not higher.  The origin of this characteristic repeating-FRB spectral structure remains unknown. \citet{cordes_lensing_2017} suggests it may arise from lensing within the host galaxy of the FRB source (for example, within a wind nebula or SNR associated with the FRB source, or an HII region within the galaxy), while  \citet{metzger_fast_2019} and \citet{margalit_constraints_2020} explain this change in peak frequency within their synchrotron maser model as a deceleration of the shock front. This makes it difficult for the ``far-away'' emission models to explain the differences in the spectra of the two components of the radio burst from \SGR{} \citep{lu_unified_2020}. The sharp cutoffs in the spectra of both bursts are  difficult to explain without contrived discontinuities in the pre-shock medium; see the Appendix of \citet{lu_unified_2020} for further details. 

In this letter, we posit that the observed variation in spectral structure between the two components of the CHIME/FRB-detected burst is due to interstellar scintillation.  In the picture we present in Section \ref{sec:model}, two separate scattering screens intervene along the line-of-sight to \SGR{}: one responsible of the observed temporal broadening of the burst and one responsible for the observed spectral structure.  The differences in the spectra of the two bursts allow us to place a lower limit on the separation between the sources of the two bursts, which can be interpreted as either motion of the emission region or a spatial separation between two independent emission regions. We discuss the implications of this limit on models for FRB-like emission from magnetars in Section \ref{sec:discussion}, and discuss ways in which our picture could be further tested in Section \ref{sec:conclusions}.  

\section{A two-screen model for the spectral properties of the two-component \SGR{} radio burst detected by CHIME/FRB} \label{sec:model}

Two-screen scattering models have recently gained traction in the study of propagation effects for extragalactic FRBs \citep{masui_dense_2015,ravi_magnetic_2016,farah_frb_2018,chime_observations_2019,macquart_spectral_2019,day_high_2020}; typically these are motivated by scattering inconsistent with expectations from the Milky Way or by incongruous temporal scatter-broadening timescales and scintillation bandwidths.  In this letter, ``scintillation'' refers specifically to spectral modulations caused by the coherent combination of radiation propagating along different paths, and ``scattering'' refers to the general occurrence of multi-path propagation, which may manifest as scintillation or temporal broadening of pulses, among several phenomena \citep{rickett_radio_1990}. For extragalactic FRBs, scintillation is typically associated with the effects of diffraction within the Milky Way, while the temporal broadening of the burst is attributed to additional scattering material associated with the circumburst medium, the ISM of the host galaxy or the halos of intervening galaxies. 

Multiple scattering screens are also inferred in the Milky Way from pulsar scattering observations \citep[e.g.][]{putney_multiple_2006}.  In some cases, such as scattering towards the Galactic Center (in observations of both Sgr\,A* and the Galactic-Center magnetar SGR\,J1745-2900), one of the scattering screens is very close to the source of emission \citep{dexter_locating_2017}.
Similar evidence for scattering near the source is seen towards the Crab pulsar, PSR\,B0531+21 \citep[e.g.][]{cordes_brightest_2004,main_mapping_2018,driessen_scattering_2019}.  Scintillation with characteristic bandwidths of 1\,MHz at 1.66\,GHz \citep{main_mapping_2018} and 2.3\,MHz at 2.33\,GHz \citep{cordes_brightest_2004} is associated with scattering in the Crab nebula due to the rapid decorrelation timescales of the scintillation pattern.

Below, we explore the viability of a similar two-screen model for \SGR{}, in which the spectral structure in the CHIME/FRB-detected burst is due to scintillation with an estimated scintillation half-width half-maximum bandwidth of 100\,MHz at 600\,MHz.  This corresponds to a bandwidth of 770\,MHz at 1\,GHz\footnote{We've assumed $\Delta \nu \propto \nu^4$.}, consistent with the combined detection of the second component by CHIME/FRB and STARE2. The observing parameters we will use are summarized in Table \ref{tbl:obsparams}.  The difference in the observed scintillation properties between the two components of the burst necessitates that the scintillation pattern upon the Earth has moved by at least the spatial scale of the scintle in the 28.97\,ms separating the two components.  We use this constraint on the velocity of the scintillation pattern to constrain the velocity of the emission region (if the same emission region is responsible for both components) or separations of two emission regions (if each component arose from a different region).

\begin{deluxetable*}{p{1cm}p{3cm}p{6cm}}
\tablecaption{Parameters measured or assumed for the CHIME/FRB bursts from \SGR{} and used throughout this work.\label{tbl:obsparams}}
\tablewidth{0pt}
\tablehead{
\colhead{Parameter} & \colhead{Value} & \colhead{Description} 
}
\startdata
      \multicolumn{3}{c}{\bf Measured Parameters}\\
      $\nu_\mathrm{ref}$ &  600\,MHz & Reference frequency for scattering and scintillation parameters\\
      $\Delta t$ & 28.97\,ms & Temporal separation between the two components\\
      $\tau_\mathrm{scat}$  & $0.759\pm0.008$\,ms & Scattering timescale fit by \protect\citet{chime_bright_2020}\\
      $\Delta \nu_\mathrm{scint}$ & 100\,MHz & Scintillation (HWHM) bandwidth estimated from the dynamic spectra of the CHIME/FRB-detected radio burst \\ 
      \multicolumn{3}{c}{\bf Derived Parameters}\\
      $\tau_\mathrm{scint}$ & 1.5\,ns & Scattering timescale corresponding to $\Delta \nu_\mathrm{scint}$\tablenotemark{a}\\
      $\Delta \nu_\mathrm{scat}$ & 200\,Hz & Scintillation bandwidth corresponding to $\tau_\mathrm{scat}$\tablenotemark{a}\\
      \multicolumn{3}{c}{\bf Fiducial Parameters}\\
      $d_\mathrm{src}$ & 10\,kpc & Fiducial distance to \SGR{}\\
      $d_\mathrm{scat}$ & 5\,kpc & Distance to the scattering screen responsible for  temporal-broadening \\
      $d_\mathrm{scint}$ & 9.984\,kpc & Distance to the scattering screen responsible for scintillation\\
      $s_\mathrm{scat}$ & 0.5 & $s_\mathrm{scat} = 1 - d_\mathrm{scat}/d_\mathrm{src}$ \\
      $s_\mathrm{scint}$ & $1.6\times 10^{-3}$ & $s_\mathrm{scint} = 1 - d_\mathrm{scint}/d_\mathrm{src}$\\
      $C_1$ & 0.957 & Equation \eqref{eqn:c1} \\
      \multicolumn{3}{c}{\bf Other Parameters}\\
      $l_d$ & & $1/e$ half-width of the spatial scintillation pattern\\
      $V_\mathrm{src,app}$ & & Apparent transverse velocity of the emission region \\
      $V_\mathrm{ISS}$ & & Velocity of the scintillation pattern on the observer plane \\
      $\Delta t_\mathrm{scint}$ & & Decoherence timescale of the scintillation pattern
\enddata
\tablenotetext{a}{Using Equation \eqref{eqn:c1}, with $C_1 = 0.957$ for a Kolmogorov phase structure function of the radiation scattered by a thin screen \citep{cordes_diffractive_1998}.}
\end{deluxetable*}

We start by considering the locales of the two scattering regions.  The NE2001 model \citep{cordes_ne2001_2002,cordes_ne2001_2003} predicts a scattering timescale of 0.048\,ms at 1\,GHz, which scales to 0.37\,ms at 600\,MHz (assuming a timescale $\tau \propto \nu^{-4}$), off only by a factor of two from the measured scattering timescale. This suggests that the temporal broadening arises in the ISM as modeled in NE2001.  In contrast, a scintillation bandwidth of 100\,MHz corresponds to a scattering timescale of 1.5\,ns, much smaller than (and therefore easily concealed by) the 0.759\,ms timescale.  Thus, we will consider below a picture in which the temporal broadening originates in the ISM while the scintillation is dominated by scattering closer to \SGR{}.  We consider the reverse scenario in Appendix \ref{sec:reversing}.

The distance to \SGR{} remains uncertain, with recent measurements of the distance to the associated supernova remnant SNR\,G57.2+0.8 including $6.6 \pm 0.7$\,kpc \citep{zhou_revisiting_2020} and $12.5 \pm 1.5$\,kpc \citep{kothes_radio_2018}.  We will adopt a fiducial distance of $d_\mathrm{src}=10$\,kpc.  Screens halfway between the observer and the source contribute most significantly to temporal broadening, so we will use a fiducial distance to the screen responsible for temporal broadening of $d_\mathrm{scat}=\frac{1}{2}d_\mathrm{src}$.  Finally, the SNR associated with \SGR{} has an angular radius of 5.5\,arcminutes.  As this SNR is a possible location of scattering near the host, we will adopt a fiducial fractional distance between \SGR{} and the screen responsible for scintillation of $s_\mathrm{scint} = 1-\frac{d_\mathrm{scint}}{d_\mathrm{src}} = 1.6 \times 10^{-3}$.  \citet{kothes_radio_2018} also identify a second arc-like feature in the GMRT 150-MHz survey radio map \citep{intema_gmrt_2017}, with an approximate angular radius of 3.6\,arcminutes (estimated from their Fig.\ 2).  This feature, which, as \citet{kothes_radio_2018} consider, could be a pulsar wind nebula (PWN)-like feature produced by the magnetar, is another possible location for scattering (with $s_\mathrm{scint}=1.0 \times 10^{-3}$).

The $1/e$ half-width of the scintillation pattern projected on the observer plane, assuming isotropic scattering in a thin screen a distance $d_\mathrm{lens}$ from the observer, is given by \citep{cordes_diffractive_1998}\footnote{This arises from $l_d = \frac{1}{\sqrt{2}\pi}\frac{c}{\nu_\mathrm{ref}d_\mathrm{src}\theta_\mathrm{rms}} \frac{1-s}{s}$.  For two rays, one unscattered and one scattered by an angle $\theta$ at a screen at a distance $d_\mathrm{src}(1-s)$ from the observer, the path length difference between the two rays changes by the wavelength $\lambda$ over a distance $\frac{c}{\nu_\mathrm{ref}d_\mathrm{src}\theta} \frac{1-s}{s}$ at the observer plane. The prefactor $\frac{1}{\sqrt{2}\pi}$ relates the $1/e$ scale of the scintle to this distance for scattering with a square-law phase structure function \citep{cordes_diffractive_1998}.} 
\begin{align}
    l_d &= \frac{1}{\nu_\mathrm{ref}} \left( \frac{c \Delta \nu_d}{2 \pi C_1} \frac{d_\mathrm{src} d_\mathrm{lens}}{d_\mathrm{src}-d_\mathrm{lens}}\right)^{1/2} \label{eqn:ld1} \;,
\end{align}
where $\nu_\mathrm{ref}$ is the reference frequency, $\Delta \nu_d$ is the half-width half-max scintillation bandwidth, and $C_1$ is a factor that depends on the spectrum of density fluctuations and the distribution of the scattering material, defined by the relationship between the scintillation bandwidth and the mean delay:\footnote{For an exponential scattering tail, the mean delay is equivalent to the $1/e$ scattering timescale.  An exponential scattering timescale is expected theoretically for a scattering screen with phase structure function index of 2, but is observed for many pulsars and FRBs.} 
\begin{equation}
    2 \pi \Delta \nu_d \tau_d = C_1 \;.\label{eqn:c1}
\end{equation}
We will use $C_1 = 0.957$ for a Kolmogorov phase structure function of the radiation scattered by a thin screen \citep{cordes_diffractive_1998}.  Adopting our fiducial values, Equation \eqref{eqn:ld1} can be written as
\begin{align}
    l_d &= 5.2\times10^7\,\mathrm{km} 
    \left(\frac{\nu_\mathrm{ref}}{600\,\mathrm{MHz}}\right)^{-1} 
    \left(\frac{\Delta \nu_d}{100\,\mathrm{MHz}}\right)^{1/2} 
    \left(\frac{d_\mathrm{src}}{10\,\mathrm{kpc}}\right)^{1/2}
    \left( \frac{(1-s)s^{-1}}{624} \right)^{1/2} \;,\label{eqn:ld2}
\end{align}
where $s=1-\frac{d_\mathrm{lens}}{d_\mathrm{src}}$.  For $s=1.6\times10^{-3}$, the spatial size of the scintillation pattern is $5.2\times10^{7}$\,km, many times the size of the Earth.\footnote{For the temporal-broadening screen, assuming $s=1/2$, we find a characteristic scintillation bandwidth of 200\,Hz and scintillation spatial scale of 2100\,km.  While this spatial scale is much smaller than the scintles induced by the screen close to the source, this bandwidth is much smaller than the CHIME/FRB frequency resolution (390.625\,kHz) \citep{chime_bright_2020} and therefore this pattern is difficult to observe without inversion of the digital filterbank.}

The differing spectra observed between the two components of the burst necessitates that, if this spectral structure is due to scintillation, the scintillation pattern has moved by at least $l_d$ within the temporal separation of the two components, $\Delta t = 28.97$\,ms, or 
\begin{equation}
    V_\mathrm{ISS} \ge \frac{l_d}{\Delta t}\;,
\end{equation}
where $V_\mathrm{ISS}$ is the velocity of the scintillation pattern in the plane of the observer.  When the scattering screen is very close to the source, this motion is related to the apparent transverse motion of the source by
\begin{equation}
    V_\mathrm{src,app} = \frac{s}{1-s} V_\mathrm{ISS}\;,
\end{equation}
allowing us to constrain the apparent transverse motion of the source to be
\begin{equation}
    V_\mathrm{src,app} \ge \frac{1}{\nu_\mathrm{ref}\Delta t} \left( \frac{c \Delta \nu_d d_\mathrm{src}}{2 \pi C_1} \frac{s}{1-s}\right)^{1/2}\;.
\end{equation}
In terms of our fiducial parameters:
\begin{equation}
    V_\mathrm{src,app} \ge  9.5\,c 
    \left(\frac{\nu_\mathrm{ref}}{600\,\mathrm{MHz}}\right)^{-1}
    \left(\frac{\Delta t}{28.97\,\mathrm{ms}}\right)^{-1} 
    \left(\frac{\Delta \nu_d}{100\,\mathrm{MHz}}\right)^{1/2}
    \left(\frac{d_\mathrm{src}}{10\,\mathrm{kpc}}\right)^{1/2}
    \left(\frac{(1-s)s^{-1}}{624}\right)^{-1/2}\;.\label{eqn:vsrc}
\end{equation}
For $s=1.6\times10^{-3}$, $V_\mathrm{src,app}\geq9.5\,c$.  We can interpret this either as superluminal motion of the emission region, or as a spatial separation of $V_\mathrm{src,app}\,\Delta t \ge 8.3\times10^4$\,km between two emission regions.  The scintillation bandwidth we use is uncertain, likely to a factor of $\sim 2$.  As we see from Equation \eqref{eqn:vsrc}, our estimate for the apparent transverse velocity of the source depends on the square root of the scintillation bandwidth and therefore, for our fiducial distances to the source and scintillation-dominating screen, our inferred velocity of the source is in the range of $6.7-13.5\,c$ or transverse separation of $5.9-11.7\times10^4$\,km.  As these variations are not sufficiently significant to affect the conclusions of this work, we consider only the fiducial value of 100\,MHz going forward. 

The diffractive scale, $r_\mathrm{diff}$, the typical spatial scale of density fluctuations at the screen, can be calculated for a scattering screen from a measurement of the scintillation bandwidth $\Delta \nu_d$ and the assumed distributions of scattering material and density fluctuations.  For isotropic Kolmogorov turbulence in a thin screen distance $d_\mathrm{lens}$, the diffractive scale is given by \citep{macquart_temporal_2013}\footnote{Note that the diffractive scale is analogous to the spatial scale of the scintillation pattern, $l_d$: $r_\mathrm{diff} = l_d s$.}:
\begin{align}
    r_\mathrm{diff} &= \frac{1}{\nu_\mathrm{ref}} \left(\frac{\Delta \nu_d\,d_\mathrm{src}\,c}{2\pi C_1}\,  s (1-s) \right)^{1/2}\label{eqn:rdiff}\\
    &= 8.3 \times 10^{9}\,\mathrm{cm} 
    \left(\frac{\nu_\mathrm{ref}}{600\,\mathrm{MHz}}\right)^{-1} 
    \left(\frac{\Delta \nu_d}{100\,\mathrm{MHz}}\right)^{1/2}
    \left(\frac{d_\mathrm{src}}{10\,\mathrm{kpc}}\right)^{1/2}
    \left(\frac{s (1-s)}{1.6\times10^{-3}}\right)^{1/2}\;.\label{eqn:rdiff-fid}
\end{align}
For $s=1.6\times10^{-3}$, $r_\mathrm{diff} = 8.3\times10^9$\,cm, larger than the typical inner scale $l_\mathrm{inner}=100$\,km in the ISM \citep{spangler_evidence_2020}.

When $r_\mathrm{diff} > l_\mathrm{inner}$, the scattering measure (SM) is related to the diffractive scale by \citep{macquart_temporal_2013}
\begin{align}
    \mathrm{SM} &= \left(\frac{1}{2}r_\mathrm{diff}\right)^{-5/3} \frac{3 \nu_\mathrm{ref}^2}{11 \pi c^2 r_e^2} \frac{\Gamma(\frac{11}{6})}{\Gamma(\frac{-11}{6})}\label{eqn:sm}\\
    &= 7.4\times10^{-5}\,\mathrm{kpc}\,\mathrm{cm}^{-20/3} \left(\frac{\nu_\mathrm{ref}}{600\,\mathrm{MHz}}\right)^{11/3}
    \left(\frac{\Delta \nu_d}{100\,\mathrm{MHz}}\right)^{-5/6}
    \left(\frac{d_\mathrm{src}}{10\,\mathrm{kpc}}\right)^{-5/6}
    \left(\frac{s(1-s)}{1.6\times10^{-3}}\right)^{-5/6}\;,
\end{align}
where $r_e$ is the classical electron radius and $\Gamma(.)$ is the Gamma function.  For $s_\mathrm{scint} = 1.6\times10^{-3}$, $\mathrm{SM} = 7.4\times10^{-5}$\,kpc m$^{-20/3}$.  This is low compared to typical measurements of the SM in the ISM using pulsar scattering, which vary between $\sim 0.01 - 100$\,kpc\,m$^{-20/3}$ \citep{bhat_multifrequency_2004}.\footnote{In contrast, for the observed scattering timescale of 0.758\,ms and assuming the responsible scattering screen is midway between the observer and the \SGR{} ($s_\mathrm{scat}=0.5$), we infer a scattering measure for material within this screen of 0.062 kpc\,m$^{-20/3}$.}  However, most studies of pulsar scattering focus on scattering timescales measured within a narrow bandwidth, and thus select against weakly scattering material.  This is compounded by the fact that few instruments are able to probe scintillation bandwidths that cover such a large fraction ($1/6^\mathrm{th}$, in this case) of the central observing frequency.  Ongoing and future wide-band pulsar studies (e.g.\ with the MWA \citep{kirsten_probing_2019}, LOFAR \citep{stappers_observing_2011}, MeerTime \citep{bailes_meertime_2018}, CHIME/Pulsar \citep{ng_pulsar_2018}, and pulsar timing with the Parkes Ultra-Wideband (UWL) receiver \citep{hobbs_uwl_2020}) will provide more insight into the prevalence of weakly-scattering material in the Milky Way. 

\begin{figure}
\centering
  \includegraphics[width=0.6\textwidth]{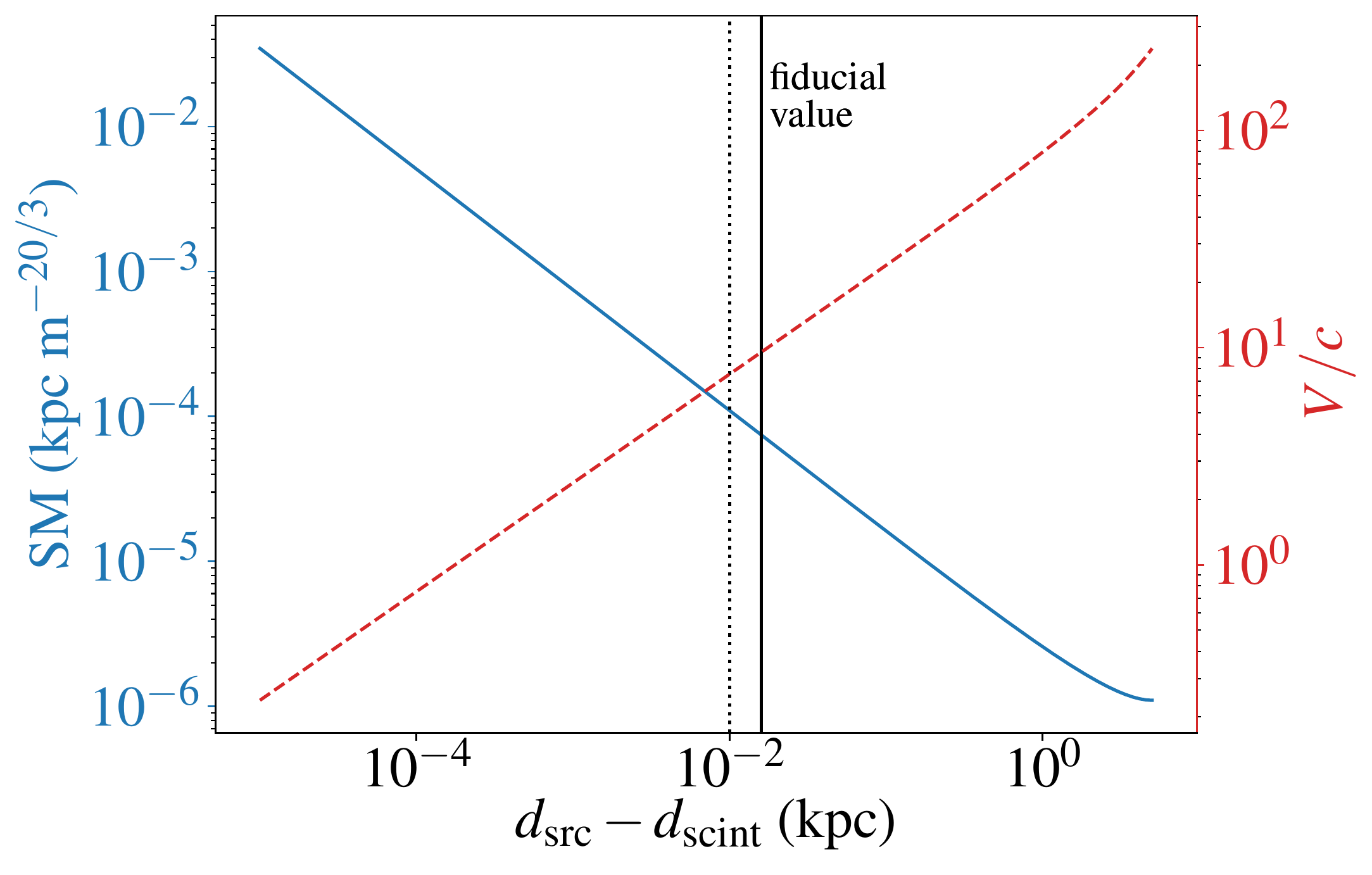}
  \caption{
  Dependence of scattering measure (SM, solid blue line, left-hand axis) and the derived lower limit on the velocity of the emission region ($V$, in units of $c$, dashed red line, right-hand axis) on the distance between \SGR{} and the scattering material responsible for scintillation.  The velocity can also be interpreted as a separation between two emission regions, each responsible for one of the observed components of the burst.  With a separation of 28.97\,ms between the components, $V/c = 1$ corresponds to a separation of 8700\,km. The fiducial separation between the screen and source of 16\,pc, for which the screen is associated with the SNR\,G57.2+0.8, is indicated with the solid black line, while the separation of 10\,pc corresponding to scattering in the putative PWN is shown as the dotted black line.  Note that as the screen is placed closer to the source, the lower limit on the velocity decreases, alleviating the constraint for superluminal motion, and the scattering measure increases, more consistent with average sight-lines through the Milky Way.  \label{fig:scintillation_parameters}}
\end{figure}

If the scattering material is in fact closer to the source, the derived SM of the material decreases while the lower limit on the speed of emission region (or on the separation between two emission regions) is relaxed, as shown in Fig.\ \ref{fig:scintillation_parameters}.  As the distance to \SGR{} is uncertain to $< 50 \%$, it has a much less significant impact on the inferred properties of the scattering material and the emission mechanism than the location of the scattering material.

\section{Discussion}\label{sec:discussion}

We have shown that the spectral features present in the bright radio burst detected by CHIME/FRB from \SGR{} can originate from scintillation due to scattering within the surrounding SNR.  If this spectral structure is indeed due to scintillation, this allows us to constrain the motion of the emission region (or the spatial separations of the sources of the two components of the burst). Here, we  consider the implications of these constraints on models for the radio emission from \SGR{}, with specific reference to the ``close-in'' \citep{lu_unified_2020,lyutikov_fast_2020} and ``far-away'' \citep{margalit_implications_2020,yuan_plasmoid_2020} classes of models. We conclude in  Section \ref{sec:conclusions} with a discussion of how future observations of \SGR{} may be able to test our interpretation of the spectral structure as scintillation. 

\subsection{Emission in the magnetosphere}\label{sec:close-in}

Several models for the generation of extragalactic FRBs from within the magnetospheres of highly magnetized NSs have been proposed \citep[e.g.][]{cordes_supergiant_2016,kumar_fast_2017,katz_coherent_2018,wadiasingh_repeating_2019,lyutikov_fast_2020,lu_unified_2020}, some of which which have recently been applied to the radio burst from \SGR{} \citep{lyutikov_fast_2020,lu_unified_2020}. \citet{lyutikov_fast_2020} simply postulate emission at $\lesssim100\,r_\mathrm{NS}$ to address equipartition considerations. \citet{lu_unified_2020} build on the model of \citet{kumar_frb_2020} to account for the joint observation of a hard X-ray burst from \SGR{} together with the radio burst. In this model, Alfv\'en waves are launched along field lines near the magnetic poles by a crustal disturbance, which dissipate upon charge starvation at heights of tens of $r_\mathrm{NS}$ (or hundreds of km). The dissipation is predicted to result in the acceleration of charge bunches with characteristic scale corresponding to the plasma-oscillation scale, resulting in a radio burst. A strength of this model is that the characteristic burst energies and frequencies naturally result from typical magnetar characteristics \citep{kumar_frb_2020}. In general, because ``close-in" models of emission in the magnetosphere cannot produce the large transverse motion (or separation) of $8.3\times10^7$\,m that we infer in our scintillation model, as this is similar to the light cylinder radius.

\citet{lu_unified_2020} conclude that the angle between the pole and the emission region must be $\sim 0.1$\,rad for the Alfv\'en waves to dissipate at a sufficient height to explain the observed radio emission.  Because the duration of each component is much smaller than the temporal separation of the components in the burst from \SGR{}, \citet{lu_unified_2020} conclude that these must be two separate emission events.  The maximum physical separation between these two events within this picture is determined by the opening angle of the emitting region, $\sim 0.01\,\mathrm{rad}$ after accounting for relativistic beaming effects. The maximum separation is then $\sim 0.01\,\mathrm{rad}\times 20\,r_\mathrm{NS} = 2\times10^{3}$\,m, four orders of magnitude less than the separation of $8.3 \times 10^7$\,m we infer from the scintillation pattern.  Even if we assume events are coming from very different regions of the magnetosphere, the height of the emission is too large by an order of magnitude. An angular separation of $\pi$ corresponds to a physical separation of $40\,r_\mathrm{NS} \approx 8\times10^6$ m.

If instead we assume naively that the two emission components arise from the same emission region within the magnetosphere of \SGR{}, the temporal and spatial separations of the two components allows us to constrain the height of the emission region (measured from the center of the NS).  In the 28.97\,ms between the two components, the magnetar has rotated only 0.056\,rad.  In this case, the emission height must be $h_{em} > 1.5 \times 10^{9}$\,m outside of the light cylinder (radius of $\sim10^{8}$\,m), to satisfy the inferred lower limit on the spatial separation of the two components.

\subsection{Synchrotron-maser model}\label{sec:far-away}

\citet{yuan_plasmoid_2020} build a model for the \SGR{} observations that posits radio emission from shocks driven by relativistic magnetospheric ejections (known as ``plasmoids'') into the surrounding electron-positron NS wind. The plasmoid properties are derived from force-free electrodynamics simulations. Continued magnetic reconnection at the ejection site amplifies the NS wind into which successive ejections are launched. Some energy is dissipated at the ejection site as hard X-rays, and the radio burst is produced through the synchrotron maser mechanism behind the decelerating shock at $\sim10^{11}$\,m from the NS. Coincidence between the radio and X-ray bursts is established partially due to the high Lorentz factor of the shock of a few hundred. In this scenario, the two components observed by CHIME/FRB from \SGR{} represent two separate plasmoid ejections. The lower limit on the spatial separation between the components that we derive under the scintillation hypothesis, $8.3 \times 10^7$\,m, is much smaller than the emission height, and this model is therefore consistent with our constraints.  

\citet{margalit_implications_2020} apply the synchrotron-maser model of \citet{metzger_fast_2019} to the burst observed from \SGR{}. In this model, the shock is driven into a medium composed of the slow baryonic tails of previous ejections, rather than into the electron-positron NS wind. Through their analysis, they infer the radius of the shock responsible for the radio burst to be $\sim 1.7\times 10^{9}$\,m at the time of emission.  Again, this is much larger than the lower limit on the spatial separation between the two emission components and so consistent with this constraint. However, \citet{margalit_constraints_2020} expect FRB emission in the surrounding medium to be suppressed for a time $\sim r_\mathrm{sh}/c$ ($\sim$ seconds) after a burst due to heating of the medium by the first shock.  \citet{babul_synchrotron_2020} find the wait times for the shocked plasma to cool sufficiently are even longer, and that the second shock must outrun the first in order for components to be explained by successive shocks.  If instead we attribute the hypothesized scintillation to motion of the emission region, we need to explain the apparent transverse velocity of $V_\mathrm{src,app} > 9.5\,c$.  \citet{margalit_implications_2020} model the shocked gas with a Lorentz factor $\Gamma \approx 24$.  In this case, an offset between the line-of-sight and the direction of motion of the emission region of only $\theta_v = 0.0086$\,rad, much smaller than the relativistic beaming angle, $\theta_b = 0.04$\,rad, is needed to produce the observed superluminal motion.  

To derive these constraints, we have assumed that motion of the emission region is marginally resolved by the screen.  This means that spatial scales smaller than $8.3\times10^7$\,m are unresolved by the scattering screen and the screen does not resolve the beamed cone of emission itself.

Generally, ``far-away'' models are developed assuming spherical symmetry in the rest frame, and the emission is relativistically beamed towards the observer. Multiple emission sites or superluminal motion are only possible if the emission regions are in fact structured on scales smaller than the relativistic-beaming cones. 


\subsection{Extragalactic FRBs}\label{sec:eg-frbs}

Like the burst from \SGR{} detected by CHIME/FRB, some extragalactic FRBs show multiple components.  When components are observed with different spectral structures, a similar analysis can be done for these FRBs.  First, we must rewrite Equation \eqref{eqn:vsrc} in terms of angular diameter distances and fiducial values more representative of the FRB population:
\begin{align}
    V_\mathrm{src,app} &\ge  130\,c 
    \left(\frac{\nu_\mathrm{ref}}{1\,\mathrm{GHz}}\right)^{-1}
    \left(\frac{\Delta t}{1\,\mathrm{ms}}\right)^{-1} 
    \left(\frac{\Delta \nu_d}{100\,\mathrm{MHz}}\right)^{1/2}
    \left(\frac{d_\mathrm{src,scint}}{10\,\mathrm{pc}}\right)^{1/2}
    \left(\frac{d_\mathrm{src}}{1\,\mathrm{Gpc}}\right)^{-1/2}
    (1+z)^{-1/2}\;,\label{eqn:vsrc_cosmo}
\end{align}
where $d_\mathrm{src,scint}$ is the angular diameter distance between the source and the screen responsible for scintillation and $z$ is the redshift of the lensing material. We have assumed that $d_\mathrm{src,scint}\ll d_\mathrm{src}$, so that the distance from the observer to the scattering screen is $d_\mathrm{scint} \approx d_\mathrm{src}$ and so that we can approximate the redshift of the screen as the redshift of the source.

Some FRBs show no evidence for scintillation (which may be masked by the instrumental resolution) \citep[e.g.,][]{ravi_pop_2019} and others show scintillation that is constant across multiple components \citep[e.g.,][]{farah_frb_2018,farah_five_2019}.  However, FRB\,190611, a two-component burst detected by ASKAP and tentatively localized to a host galaxy at $z=0.378$ \citep{macquart_census_2020} shows qualitative similarities with the picture described here \citep{day_high_2020}.  The two components separated by $\sim 1$\,ms show evidence of scintillation from the Milky Way ISM (consistent between the two components) in addition to an overall envelope which has a central frequency 48\,MHz higher for the latter component \citep{day_high_2020}.  Estimating $\Delta \nu = 100$\,MHz at 1250\,MHz, assuming the screen, like in the case of \SGR{}, is 16\,pc from the source\footnote{Here, we are essentially assuming that the source of FRB\,190611 is, like \SGR{}, a magnetar embedded in a SNR with radius 16\,pc.}, and using the Planck 2015 cosmological parameters \citep{planck_2015} to calculate the angular diameter distance to the source, we find $l_d = 3.2 \times 10^{13}$\,km and $V_\mathrm{src,app} > 160\,c$, implying a highly relativistic emission region, with $\Gamma \gtrsim V_\mathrm{src,app}/c$ or a separation between two emission regions of $4.7\times10^7$\,m.  Such high Lorentz factors are predicted by \citet{beloborodov_blast_2019} and \citet{margalit_constraints_2020} for extragalactic FRBs.

FRB\,190611 shows time-varying polarization properties as well as an apparent change in dispersion measure between the two components \citep{day_high_2020}.  The apparent variations in dispersion measure, Faraday rotation measure, and polarization properties between the components of some FRBs \citep[e.g.][]{cho_spectro_2020,day_high_2020} may be interpreted as emission observed along different sightlines through a dense, magnetized, trans-relativistic plasma \citep{vedantham_faraday_2019}, possibly with several radial magnetic-field reversals \citep{gruzinov_conversion_2019}. This scenario is consistent with our picture of significant spatial separations between emission sites for multiple-component FRBs. The apparent dispersion-measure change may also be intrinsic to the emission mechanism (e.g.\ due to slight offsets in the times of emission at different frequencies); similar phenomena are sometimes observed from FRB\,121102 \citep{hessels_frb_2019}.  

\section{Conclusions}\label{sec:conclusions}

Scintillation is a viable explanation for the spectral differences between the two components of the bright radio burst detected by CHIME/FRB from \SGR{}. If this explanation is correct in practise, and using a model in which temporal broadening is dominated by a screen in the Milky Way ISM while scintillation is dominated by a screen in the SNR associated with \SGR{}, we place a lower limit on the separation between the emission regions (or the motion of the emission region between the two bursts) of $8.3\times10^4$\,km (or $9.5\,c$).  This separation is inconsistent with ``close-in'' models of emission within $\sim100\,r_\mathrm{NS}$ of the magnetar surface \citep{lyutikov_fast_2020,lu_unified_2020}, but can be explained by ``far-away'' emission models that posit radio emission from well beyond the magnetosphere \citep{margalit_implications_2020,yuan_plasmoid_2020}.  The observed difference in the spectra between the two components could instead be intrinsic to the emission mechanism - variations on such timescales are expected in the ``close-in" model of \citet{lu_unified_2020}.  Further observations are therefore vital to test our model of scintillation and determine the weight of the implications discussed in this letter.

In the case of a single moving emission region responsible for the multiple components in the \SGR{} burst, the beaming angle must be greater than the angle between the emission region velocity and the line-of-sight in order for the burst to be observed. This implies that the apparent transverse velocity must be $< \Gamma\,c$. If the emission region were moving at this speed, the decoherence timescale of the scintillation pattern, $\Delta t_\mathrm{scint}$, would be decreased according to $\Delta t_\mathrm{scint} \propto V_\mathrm{src,app}^{-1}$.  This allows us to infer a minimum decoherence time of 
\begin{align}
    \Delta t_\mathrm{scint,min} &= \frac{l_d}{V_\mathrm{src,max}}\frac{s}{1-s}\\
                          &= 2.8\,\mathrm{ms} \left(\frac{\Gamma}{100}\right)^{-1}
    \left(\frac{\nu_\mathrm{ref}}{600\,\mathrm{MHz}}\right)^{-1} 
    \left(\frac{\Delta \nu_d}{100\,\mathrm{MHz}}\right)^{1/2} 
    \left(\frac{d_\mathrm{src}}{10\,\mathrm{kpc}}\right)^{1/2}
    \left( \frac{(1-s)^{-1}s}{1.6\times10^{-3}} \right)^{1/2} \;.\label{eqn:decoherence}
\end{align}
Components separated by less than this timescale will show the same spectral structure.  If there are additional factors that set the beaming cone to be smaller than this relativistic limit, the decoherence timescale may be smaller.  This decoherence timescale analysis also applies to scintillation of extragalactic FRBs, and could be applied to verify our model of scintillation as the origin of spectral structure in cases like FRB\,190611.  Generically, the scintillation bandwidth is expected to increase at higher frequencies as $\Delta \nu_d \propto \nu^{\alpha}$.  Typically, this follows a power-law relation, $\Delta \nu_d \propto \nu^\alpha$, where $\alpha=4$ for a square power-law distribution of density fluctuations and $\alpha=4.4$ for a Kolmogorov distribution.  Pulsar observations typically find values of $\alpha$ between 1.5 and 4.5 \citep[e.g.][]{bhat_multifrequency_2004,geyer_scattering_2017,kirsten_probing_2019}.  Adopting $\alpha=4$ and assuming no dependence of $\Gamma$ on frequency, we then expect this decoherence timescale to scale as $\Delta t_\mathrm{scint,min}\propto \nu$.  In general, ``far-away'' models predict a weak dependence of the Lorentz factor of the relativistic shock on the radio-burst energy; however, the frequency of the observed radiation scales with the Lorentz factor and the local electron gyroradius, which in turn depends on the local pre-shock magnetic field \citep{beloborodov_blast_2019,metzger_fast_2019}.

Ultra-wideband or simultaneous multi-band observations will also allow tests of this scintillation model, by characterizing the frequency-dependence of the decoherence bandwidth, which we predict to scale as $\Delta \nu_d \propto \nu^\alpha$ with $\alpha \approx 4$, as discussed above.  Similar studies of the original repeating FRB source, FRB\,121102, have characterized the observed drift of components to lower frequencies at later times \citep[e.g.][]{hessels_frb_2019,chime_detection_2019}. \citet{caleb_simultaneous_2020} fit the drift rate as a function of frequency using a linear model, $\frac{\mathrm{d}\nu}{\mathrm{d}t} \propto \nu$.  In our model, this drift rate is analogous to the ratio $\Delta \nu_d / \Delta t_\mathrm{scint} \propto \nu^3$, for which we expect a frequency dependence inconsistent with that in FRB\,121102.  Our model also cannot explain the lack of observations of upward-drifting components in FRB\,121102 - we expect to see both upward and downward frequency drifts of the components due to scintillation.

Like the decoherence time, the spatial coherence scale of the scintillation pattern scales as $l_d \propto \nu$.
While the spatial scale thus decreases at lower frequencies, we would have to observe at frequencies $\lesssim 100$\,MHz, below the lowest-frequency FRB detections (300\,MHz) to date \citep{pilia_lowest_2020,chawla_detection_2020}, to resolve the spatial scintillation pattern we predict for \SGR{} using stations across the Earth.  At $100$\,MHz, the scintillation bandwidth is predicted to be $\sim 24$\,kHz, requiring observations with fine spectral resolution.  However, the detection of a burst from \SGR{} showing different spectral patterns in two stations across the Earth would provide an unequivocal confirmation that this spectral pattern arises from scintillation.  Because of the much further distances of extragalactic FRBs, we expect the scintillation patterns from this model applied to extragalactic sources to have much greater spatial scales ($\sim 10^{13}$\,km).  Therefore, \SGR{} provides a unique opportunity to test this model of scintillation producing differences in FRB spectra on short timescales using multi-station observations. 

Finally, all coherent emission from \SGR{} is subject to the same scattering effects as the radio burst we've analyzed here. If the magnetar exhibits proper motion similar to typical transverse pulsar velocities, on the order of $\sim 100$\,km\,s$^{-1}$, the magnetar will traverse $8.3\times10^4$\,km in $\sim$ 83\,s.  Some magnetars exhibit pulsed radio emission \citep[e.g.][]{camilo_variable_2007,camilo_magnetar_2008,levin_radio_2010,levin_radio_2012}, including at frequencies below $150$\,MHz \citep{malofeev_new_2012,glushak_discovery_2014}. If \SGR{} exhibits similar emission \citep[which has already been tentatively detected;][]{zhang_highly_2020,burgay_atel_2020}, we expect evidence of scintillation from within the SNR in the form of a spectral pattern that decoheres between bursts with temporal separations on the order of minutes at a reference frequency of 600\,MHz (or decoheres between stations at frequencies $\lesssim 100$\,MHz as described above).  

Future monitoring of \SGR{} at radio frequencies is therefore crucial for confirming our model of scintillation towards \SGR{} and the resulting constraints we have placed on the emission mechanism for the bright radio burst observed from \SGR{}. If ``far-away'' emission models indeed apply to FRBs at extragalactic distances, our analysis suggests that scintillation enables us to resolve the emission regions of multiple-component or long-duration bursts with sub-nanoarcsecond angular resolution. 




\acknowledgments

We thank Christopher Bochenek, Casey Law, and Wenbin Lu for helpful feedback on early drafts of this work.  This research was supported by the National Science Foundation under grant AST-1836018.

\software{astropy \citep{astropy_astropy_2013,astropy_astropy_2018},  
          pyne2001 (\pyne{}),
          NE2001 model \citep{cordes_ne2001_2002,cordes_ne2001_2003}
          }

\appendix

\section{Reversing the order of the screens}\label{sec:reversing}

In the main body of this paper, we assumed that the screen responsible for temporal scattering was closer to the observer than the screen responsible for scintillation, motivated by the consistency of the scattering tail timescale with expectations from the NE2001 model of scattering and dispersion in the Milky Way \citep{cordes_ne2001_2002,cordes_ne2001_2003}.  In this section, we consider the impact of reversing the order of the screens. We first place the screen responsible for scintillation half-way to \SGR{}, $d_\mathrm{scint} = 0.5 \,d_\mathrm{src}$.  Using Equation \eqref{eqn:vsrc}, this would require $V_\mathrm{src,app} = 240\,c$ or a separation of emission regions $> 2.1\times10^6$\,km.

In order for scintillation from the screen closer to the observer to be present in this scenario, the scatter-broadened image of the source on the further screen must be unresolved to the closer screen.  The resolution of the screen responsible for the observed scintillation is given by (again assuming isotropic Kolmogorov turbulence in a thin screen)
\begin{align}
    \theta_\mathrm{res} &= \frac{c}{\nu_\mathrm{obs}}\frac{1}{\theta_\mathrm{scint} d_\mathrm{scint}}\\
    &= \frac{1}{\nu_\mathrm{obs}} \left(\frac{\pi c \Delta \nu_d}{ C_1}\frac{1}{d_\mathrm{src}\,s_\mathrm{scint}\,(1-s_\mathrm{scint})}\right)^{1/2}\\
    &= 6.1\,\mu\mathrm{as}
    \left(\frac{\nu_\mathrm{obs}}{600\,\mathrm{MHz}}\right)^{-1}
    \left(\frac{\Delta \nu_d}{100\,\mathrm{MHz}}\right)^{1/2}
    \left(\frac{d_\mathrm{src}}{10\,\mathrm{kpc}}\right)^{-1/2}
    s_\mathrm{scint}^{-1/2}\left(1-s_\mathrm{scint}\right)^{-1/2}\;,
\end{align}
where $\theta_\mathrm{scint}$ is the angular size of the scatter-broadened image on the screen responsible for the observed scintillation (the screen closer to the observer).  We have made use of the relation between the scattering timescale and the angular size of the image,
\begin{equation}
    \tau = \frac{1-s}{s} \frac{\theta^2 d_\mathrm{src}}{2c}\;,
\end{equation}
as well as equation \eqref{eqn:c1}.  When $s_\mathrm{scint}=0.5$, the resolution of the scintillation-dominating screen is 12\,$\mu$as.  In order for the scattering screen to not broaden the image of the source beyond this resolution,
\begin{equation}
    s_\mathrm{scat} < \left(1+\frac{2c\tau_\mathrm{scat}}{d_\mathrm{src} \theta_\mathrm{res}^2}\right)^{-1}\;.
\end{equation}
For $\theta_\mathrm{res} = 12\,\mu$as and $d_\mathrm{src} = 10$\,kpc, $s_\mathrm{scat} < 2.4\times10^{-6}$, 
or $d_\mathrm{src}-d_\mathrm{scat} < 2.4 \times 10^{-2}$\,pc.  It is difficult to motivate scattering so close to \SGR{}.  Even if the scintillation-dominating screen is placed at the edge of the Local Bubble, $d_\mathrm{scint} = 46$\,pc and the resolution of this screen decreases to 90\,$\mu$as, the temporal broadening-dominating screen must still be $<1.3$\,pc from the source, while the spectral structure variation between the two components would necessitate an apparent transverse velocity $> 3\times10^8\,c$ of the emission region.  This derived unreasonable proximity of the screen to the source further supports our assumption, used throughout the main text, that the scintillation-dominating screen is further from the observer than temporal broadening-dominating screen.

\bibliography{main}{}
\bibliographystyle{aasjournal}

\end{document}